**Did Uranus' regular moons form via a rocky giant impactor?**

Jason Man Yin Woo[1,2,3], Christian Reinhardt[1], Marco Cilibrasi[1,4], Alice Chau[1], Ravit Helled[1], Joachim Stadel[1]

[1]Institute for Computational Science, University of Zürich, Winterthurerstrasse 190, 8057 Zürich, Switzerland
[2]Institut für Planetologie, University of Münster, Wilhelm-Klemm-Straße 10, 48149, Germany
[3]Laboratoire Lagrange, Université Côte d'Azur, CNRS, Observatoire de la Côte d'Azur, Boulevard de l'Observatoire, 06304 Nice Cedex 4, France
[4]Department of Physics, ETH Zurich, Wolfgang-Pauli-Strasse 27, 8093 Zürich, Switzerland

**Abstract**

The formation of Uranus' regular moons has been suggested to be linked to the origin of its enormous spin axial tilt (~98°). A giant impact between proto-Uranus and a 2-3 $M_{Earth}$ impactor could lead to a large tilt and to the formation of an impact generated disc, where prograde and circular satellites are accreted. The most intriguing features of the current regular Uranian satellite system is that it possesses a positive trend in the mass-distance distribution and likely also in the bulk density, implying that viscous spreading of the disc after the giant impact plays a crucial role in shaping the architecture of the final system. In this paper, we investigate the formation of Uranus' satellites by combining results of SPH simulations for the giant impact, a 1D semi-analytic disc model for viscous spreading of the post-impact disc, and N-body simulations for the assembly of satellites from a disc of moonlets. Assuming the condensed rock (i.e., silicate) remains small and available to stick onto the relatively rapid growing condensed water-ice, we find that the best case in reproducing the observed mass and bulk composition of Uranus' satellite system is a pure-rocky impactor with 3 $M_{Earth}$ colliding with the young Uranus with an impact parameter $b$ = 0.75. Such an oblique collision could also naturally explain Uranus' large tilt and possibly, its low internal heat flux. The giant impact scenario can naturally explain the key features of Uranus and its regular moons. We therefore suggest that the Uranian satellite system formed as a result of an impact rather than from a circumplanetary disc. Our results also suggest that objects beyond the water snow-line could be dominated by rocky objects similar to Pluto and Triton. Future missions to Uranus and its satellite system would further constrain the properties of Uranus and its moons and provide further insight on their formation processes.

1. Introduction

The gas giants Jupiter and Saturn are thought to form their regular satellites from a circumplanetary disc which is generated from gas and dust directly accreted from the protoplanetary disc (e.g. Canup and Ward, 2006; Mosqueira and Estrada, 2003) during the formation of the gas giants. While a similar scenario has also been proposed for the formation of the satellite systems of the less massive ice giants (Canup and Ward, 2006; Szulágyi et al., 2018), it is currently still unclear whether Uranus and Neptune can acquire a circumplanetary disc during their formation since these planets are not expected to undergo runaway gas accretion (Helled et al., 2020 and references therein).

Uranus has a remarkable spin-axial tilt of ~98° relative to its orbital plane. It has been hypothesised that Uranus' large obliquity was caused by a giant impact late in its formation (e.g. Safronov, 1966; Slattery et al., 1992). Recent high-resolution smoothed particle hydrodynamics (SPH) simulations show that such a late giant impact can not only explain Uranus' current axial tilt, spin rate, and low thermal flux (Fortney et al., 2011; Nettelmann et al., 2013; Pearl and Conrath, 1991), but also to deposit enough debris into a circumplanetary orbit after the collision that can lead to the formation of its regular satellites system (Kegerreis et al., 2018; Kurosaki and Inutsuka, 2018; Reinhardt et al., 2020).

The mass of the Uranian satellite system is mostly concentrated in the largest five members with prograde and nearly circular orbits: Miranda, Ariel, Umbriel, Titania and Oberon, from the closest to farthest. Their masses range from ~$10^{-6}$ to ~$10^{-4}$ $M_U$ and their total masses are $1.05 \times 10^{-4}$ $M_U$, where $M_U = 8.7 \times 10^{25}$ kg is Uranus' mass. In this study, we will focus on studying the formation process of these five most massive satellites of Uranus from an impact generated disc. An intriguing feature of the Uranian satellite system is the positive trend of the mass-distance distribution, in which the innermost Miranda is an order of magnitude less massive than the intermediate Ariel and Umbriel; and the outermost Titania and Oberon are the most massive. Recent N-body simulations have failed to obtain such a positive trend of the mass-distance distribution with the initial conditions following the non-spreaded impact generated disc, which are very compact in the innermost region, right after the giant impact (Ishizawa et al., 2019).

To overcome this problem, the thermal and viscous spreading of the impact generated disc should be considered in order to explain the positive trend of the mass-distance distribution. It has been suggested that all the satellites or proto-satellites of Uranus are formed at/near the Roche radius as the compact disc (or ring) undergoes viscous spreading (Crida and Charnoz, 2012). In this scenario the more massive satellites and their building blocks form first at the Roche radius, followed by the smaller satellites as they migrate outward due to Uranus' tidal force. This model naturally explains the positive trend of the mass-distance distribution of the current system. However, recent SPH simulations show that in some cases more than half of the impact generated disc's mass is distributed outside the Roche radius of the planets right after the giant impact (Reinhardt et al., 2020), and this material outside Roche radius can also contribute to the formation of satellites. Hence, an alternative scenario is that the satellites of Uranus form more or less in-situ after the disc viscously spreaded.

In a later study of including viscous spreading into the impact generated disc, Ida et al. (2020) successfully reproduced the positive trend of the mass-distance distribution by the combination of a 1D viscous diffusion equation (Hartmann et al., 1998) and analytical models for dust condensation and evolution. According to their analytical solution, more ice (in this study, we refer to water ice only unless specified) condensed in the outer region as the disc spreads and cools, leaving the inner disc depleted in water ice as water vapour is lost to Uranus when the disc spreads. As a result, the surface density of the condensed ice has a positive gradient which can explain the more massive satellites in the outer region.

In addition to the positive trend of the mass-distance distribution, another peculiar feature of the Uranian satellite system is that the satellites are thought to be composed of roughly half rock and half ice, with the innermost satellite Miranda being the most icy one (Jacobson et al., 1992). So far no studies have attempted to combine impact simulation results with viscous spreading of the impact generated disc to explain the bulk composition of the Uranian satellites. Ida et al. (2020) suggest that their model can explain the roughly 1 to 1 rock-to-ice ratio of the satellites, although they considered only the icy component of the disc, since early condensed silicate particles remain small and could potentially stick to later condensed ice. However, it is clear that further investigations are required to verify this claim.

In this paper, we perform a follow-up study of Reinhardt et al. (2020) and investigate which SPH giant impact results can reproduce both the positive trend of the mass-distance distribution, as well as the roughly 1 to 1 rock-to-ice ratio of Uranian regular satellites. We investigate only the oblique collisions with impact parameter $b ≥ 0.6$, which have the potential to reproduce the large tilt and low internal heat flux of Uranus (Reinhardt et al., 2020; hereafter R20). Our study consists of three parts: SPH results (R20; Section 2.1), a 1D semi-analytical disc evolution model following Cilibrasi et al. (2018, 2021) to viciously evolve the disc (Section 2.2) and finally the GPU-based N-body simulations (Grimm and Stadel, 2014) to form the satellites from a disc of moonlets generated according to the results of the 1D semi-analytical disc evolution model (Section 2.3). Unlike Ida et al. (2020), we do not rely solely on analytical theory, but also on results from various numerical models to explain Uranus satellites formation. We also consider the rocky component of the impactor in order to compute the bulk composition of the satellites, which has not been done in the previous studies.

We present our results of the disc evolution model and N-body simulation in Section 3. We discuss the limitations and implications of our results in Section 4. We conclude our results in the final section.

## 2. Method

### 2.1. SPH simulations

As in R20 we use the SPH code Gasoline (Wadsley et al., 2004) with numerical improvements for planetary collisions (Reinhardt and Stadel, 2017; R20) to simulate the giant impacts. The particle representations of the planets prior to the impacts are obtained using ballic (Reinhardt and Stadel, 2017; R20). R20 assumed a three layer structure (e.g. Nettelmann et al., 2013) for Uranus: a rocky core, an inner water and outer H-He envelope (see R20 for details). We re-simulated the most promising cases found in R20 with more realistic and thermodynamically consistent equations of state (EOS). We use ANEOS (Meier et al., 2021; Thompson and Lauson, 1974), a commonly-used EOS for impact simulations due to its large range of validity and the thermodynamically consistent treatment of phase transitions and phase mixtures (Benz et al., 1989), to model the heavy elements. For rock we use dunite (Benz et al., 1989) and water ice (Mordasini, 2020) for water. The H-He envelope is modelled as a linear mixture between H-REOS.3 and He-REOS.3 (Becker et al., 2014) with a He mass fraction of Y=0.275 using the additive volume law. Details on how the choice of EOS affects the impact outcome for given impact conditions can be found in Appendix A2.

In R20 an impact generated circumplanetary disc was considered a potential proto-satellite disc if (i) it contains at least the total mass of Uranus' regular satellites in rock and ice, (ii) extends beyond the distance of Oberon and contains at least one Oberon mass (~3.5 × $10^{-5}$ $M_U$) in rock and ice beyond this distance, and (iii) has a minimum rock mass of half the total current satellites' mass. Several of the impacts that substantially altered Uranus' tilt fulfill these requirements. Table 1 in the Appendix A1 shows the disc's parameters at the end of the SPH simulations that we consider in this study. In general, the total mass (including H-He, rock and water ice) of the impact generated discs shortly after the giant impact is one to two orders of magnitude more massive than the current Uranian satellites system (~$10^{-4}$ $M_U$). The subsequent viscous spreading of the disc would deplete > 99% of the disc mass, leaving the disc with the appropriate amount of rock and ice mass to form the satellite system.

## 2.2. 1D semi-analytical disc evolution

The SPH simulations provide the total H-He, ice, and rock mass that is ejected into the disc. We then study the disc's evolution by using a 1D semi-analytical approach, similar to the one already implemented for satellite population synthesis by Cilibrasi et al. (2018, 2021). First, we rely on the analytical solution of the 1D viscous disc equation for the gas component (Pringle, 1981):

$$\frac{\partial \Sigma_g}{\partial t_{disc}} = \frac{1}{r}\frac{\partial}{\partial r}\left[3r^{1/2}\frac{\partial}{\partial r}\left(\Sigma_g v r^{1/2}\right)\right] \quad (1)$$

provided by Ida et al. (2020), where $\Sigma_g$ is the surface density of the gas, $v = \alpha c_s^2 \Omega^{-1}$, where $c_s$ is the sound's speed, $\Omega$ is the orbital frequency of the disc gas, and $\alpha = 10^{-3}$ is a constant parameter that represents the turbulence strength. At $t_{disc} = 0$ years, all the materials are supposed to be vapourized and distributed as:

$$\Sigma_g(r) = \Sigma_{g0}\left(\frac{r}{R_U}\right)^{-3/4}\exp\left[-\left(\frac{2r}{3r_{disc}}\right)^{5/4}\right], \quad (2)$$

where $R_U$ is Uranus' radius and $\Sigma_{g0}$ is chosen so that the total mass agrees with the value provided by the SPH simulations. $r_{disc}$ is the mean radius of impact generated disc before viscous spreading, taken from the SPH simulations and is defined as

$$r_{disc} = \left(\frac{J_{disc}/M_{disc}}{R_U^2 \Omega_U}\right)^2 R_U, \quad (3)$$

where $J_{disc}$ is the total angular momentum of the disc, $M_{disc}$ is the total mass of the disc, and $\Omega_U$ is the angular velocity of the disc at $r = R_U$ (Ida et al., 2020). This analytic solution agrees well with the surface density calculated from the SPH particles which in turn approximately follows a power-law with $\Sigma_g \sim r^{-3.25}$ similar to Ida et al. (2020). As in Ida et al. (2020) smaller differences will vanish during the later evolution due to viscous spreading of the disc and not affect our conclusions.

As long as a component (H-He, water or rock) is in the vapour state, its viscous evolution is computed following the analytical solution of Ida et al. (2020) (Eq. 1), including the temperature calculations (Eq. 2 of Ida et al. (2020)). When the temperature reaches a value below 2000 K (Melosh, 2007), we assume that all the rock in that location condensed, and therefore the mass of rocks is shifted from the vapourised rock density profile to the solid rock density profile. This also occurs for the water when the temperature drops below 240 K (Ida et al., 2020), with vapourized water density profile being shifted to an ice density profile. We track a total of three vapour density profiles that follow an analytical evolution and two solid density profiles (rock and ice), which are evolved with a semi-analytical procedure that accounts for the interaction with the vapour.

To calculate the radial drift of the condensed solids, we first calculate the coupling between the condensed solid and vapour component via the Stokes number:

$$St = \frac{\pi}{2} \frac{a \rho_s}{\Sigma_g}, \tag{4}$$

where $a$ is the size of the solid particles and $\rho_s$ is their density. If the Stokes number approaches zero, the solid particles are well-coupled to the gas and evolve radially with the same analytical solution as in Eq. 1 and Eq. 2. If the Stokes number is high, the particles start to decouple from the gas and their radial speed is given by friction via

$$v_{r,\text{fric}} = H^2 \frac{St}{1+St^2} v_K, \tag{5}$$

where $v_K$ is the Keplerian velocity, and $H \sim c_s/v_K$ is the aspect ratio of the disc. As expected, the radial velocity becomes zero when the solid particles get very large (together with *St*) and completely decouple from the gas. To consider both the radial velocities given by friction $v_{r,frit}$ and diffusion given by the coupling with the gas $v_{r,coup}$, the total radial velocity of solids is set to be

$$v_r = v_{r,\text{fric}} + \frac{v_{r,\text{coup}}}{1+St^2} = \frac{H^2 St\, v_K + v_{r,\text{coup}}}{1+St^2}. \tag{6}$$

where $v_{r,coup}$ is derived by solving the radial mass-conservation equation (see Eq. 7) with Eq. 1 (Birnstiel et al., 2010; Youdin and Lithwick, 2007). In this formalism, the coupling velocity models the advection of solids caused by the interaction with the gas, while the friction velocity models only the radial drift given by the loss of angular momentum due to the friction with the gas. This radial velocity $v_r$ is then applied to calculate the solid flux needed to solve a 1D advection equation in the form of the mass-conservation equation:

$$\frac{\partial \Sigma_s}{\partial t_{\text{disc}}} + \frac{1}{r} \frac{\partial (r \Sigma_s v_r)}{\partial r} = 0, \tag{7}$$

where $\Sigma_s$ is the surface density of the given solid species (either silicate rock or water ice).

At this stage, we neglected the effect of condensed rock and ice diffusion. The diffusion equation is practically slower to solve as the timestep $\Delta t$ has to be $\Delta t \sim \Delta r^2$ in order to guarantee the stability of the finite difference scheme (while $\Delta t \sim \Delta r$ for advection). As we will show in Section 4.4, the effect of diffusion on the final configuration of satellites (in mass, position and density) turns out to be negligible. We thus decided to exclude it and be able to explore a wider range of initial conditions and parameters.

We perform semi-analytical simulations with the 1D disc evolution model for each set of parameters provided by SPH simulations (see Appendix A1 and Table 1) for 10,000 years, in order to generate the rock and ice surface density profiles to be used by the N-body simulations.

### 2.3. N-body simulations

We next apply the results at 10,000 years of the 1D disc evolution model (see the previous section) to generate the initial conditions of the moonlets for N-body simulations. We ensure that the disc has spread and cooled sufficiently so that moonlets can be formed when the solid (condensed ice and rock) to gas ratio of the local region is larger than one (Cilibrasi et al., 2018; Drążkowska and Dullemond, 2014). The solid surface density of the moonlets follows the surface density of the condensed ice and rock at 10,000 years computed from the 1D disc evolution model. Only regions beyond the synchronised radius of Uranus (~3.3 $R_U$) are considered, since within the synchronised radius the satellite formed will be driven towards the planet by the tidal force.

The bulk density of the moonlets are computed based on the local condensed ice to rock ratio. We generate 20,000 equal mass moonlets for each simulation. The initial mass of each moonlet varies between each simulation, depending on the total mass of the condensed solids. The eccentricities and inclinations of the planetesimals are uniformly random in the range $0 < e < 0.01$ and $0° < i < 0.5°$. The nodal angles and mean anomalies of the moonlets are distributed randomly. We then perform the simulations with Uranus and a disc of planetesimals with the GPU-based N-body code *GENGA* (Grimm and Stadel, 2014), treating the full self-gravity between the moonlets. Simulations are performed for 5000 years with a timestep of 0.01 days. The gas disc is neglected since the H-He gas has depleted sufficiently at 10,000 years and thus the effect of Type-I migration on the moonlets is weak, as has been shown by Ida et al. (2020).

### 3. Results

We present the results of the most successful case which reproduces both the mass-distance distribution and the bulk composition of the satellites. The study of how different parameters and assumptions affect our results are presented in Section 4. We find that the characteristics of the current satellites are best reproduced by the following case:
(1) a 3 $M_{Earth}$ fully rocky impactor colliding with the proto-Uranus with $b = 0.75$ and $v_{imp} = 18.2$ kms$^{-1}$;
(2) condensed ice grows rapidly and decouples from the gas quickly ($St \rightarrow \infty$);
(3) condensed rock remains micrometre size and couples well with the gas ($St \rightarrow 0$);
(4) condensed rock sticks to the condensed ice in the same orbits.

Points (2) to (4) have been supported by the analytic theory of Ida et al. (2020) and the references therein, as icy particles are stickier than silicate particles (Blum and Wurm, 2008; Gundlach and Blum, 2014; Poppe et al., 2000).

The initial parameters of the disc are listed in Table 1 in Appendix A1 (Disc A2A). Fig. 1 shows the snapshots of Uranus and the impact generated disc from the SPH simulations 3.32 days after the impact. The total mass of the disc is ~5 ×$10^{-2}$ $M_U$, which is more than two orders of magnitude larger than the current satellite system's mass. The total mass of rock and ice comprises about half of the initial disc mass.

Right after the giant impact, the disc is hot and compact with about half of the disc's mass bounded within $r_{disc}$ = 2.55 $R_U$ (the red circle on the left panel and the white circle on the right panel of Fig. 1). The temperature of most parts of the disc are nearly 10,000 K, implying that most of the rock and ice are still in vapour form. Different materials (shown as different colour SPH particles in Fig. 1) are homogeneously distributed throughout the disc, indicating that the disc is well mixed at this point in time.

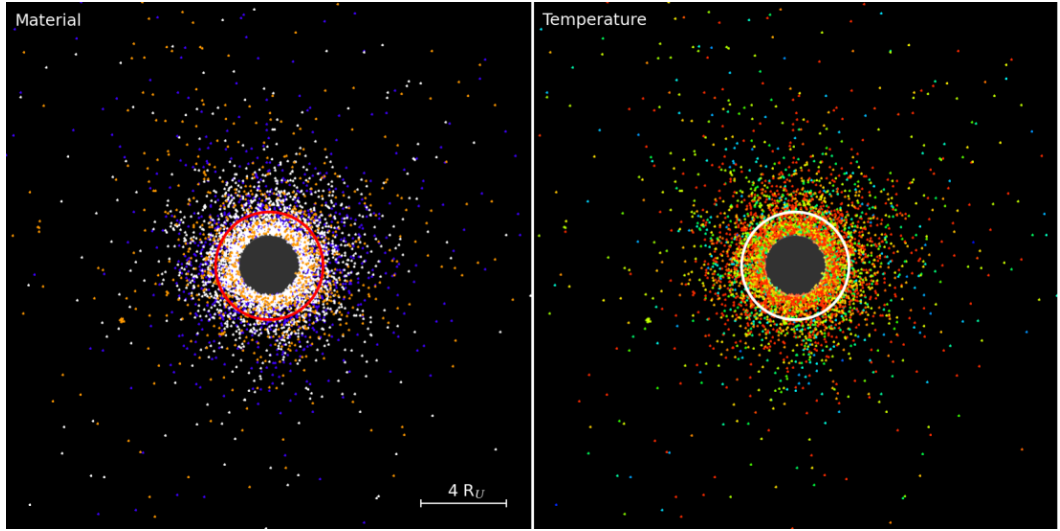

Fig. 1 - Snapshots of the impact generated disc from the SPH simulations 3.32 days after the impact (Disc A2A; Table 1). The left panel shows the material (rock: orange, ice: blue and H-He: white) and the right one temperature from 2 x $10^3$ (blue) to $10^4$ K (red). The size of one snapshot is 25 $R_U$ (about the orbit of Umbriel). The red circle on the left panel and the white circle on the right panel show the mean disc radius $r_{disc}$ = 2.55 $R_U$ within which about half of the disc mass is contained. The orbiting material consists of 5% of the total bound mass and is resolved with 5768 SPH particles. This disc is initially very hot ($T$ ~ 10,000 K) and the different materials are well mixed and homogeneously distributed throughout the disc. All of the rock in orbit comes from the impactor.

### 3.1. Viscous spreading of the disc

Next, we evolve the impact generated disc with the 1D semi-analytical disc evolution model described in Section 2.2. Fig. 2 shows the time evolution of the surface density and the temperature of the impact generated disc. The initial surface density of the disc is modeled according to Eq. 2. At $t_{disc}$ = 0 years, silicate particles (i.e. rock) condense beyond ~7 $R_U$ as the temperature beyond this distance falls below the silicate condensation temperature ($T_{rock}$ ~ 2000 K; Melosh, 2007). Icy particles condense at distances beyond ~15 $R_U$ from Uranus due to the lower condensation temperature of water ($T_{ice}$ ~ 240 K; Ida et al., 2020). Within a hundred years, the gas disc spreads beyond 100 $R_U$ and the temperature of its inner region ($r$ < 20 $R_U$) drops below $T_{rock}$, but remains higher than $T_{ice}$. Hence within 20 $R_U$ only silicate particles condense but not the icy material. As silicate material remains small (~1 μm), they couple well with the spreading gas ($St \to 0$) and thus adopt a similar surface density profile to the gas. The icy particles, on the other hand, remain in a similar surface density profile to the beginning as they grow quickly to km-sized moonlets and decoupled with the gas ($St \to \infty$).

From 100 to 1000 years, the gas surface density profile remains similar, but the disc continues to disperse and cool down. The temperature in between 8 to 20 $R_U$ now also drops below the ice condensation temperature and thus leads to condensation of icy particles. Similar to the results from the analytical study of Ida et al. (2020), we observe a positive gradient for the surface density profile of the condensed ice starting to develop. Due to the continuous viscous spreading of the disc, gas, including uncondensed ice, in the innermost region falls onto Uranus by losing angular momentum. Therefore, the outer region of the disc is more enriched in icy material than the inner region (Ida et al., 2020).

From 1000 to 10,000 years, the ice condensation front continues to push inward as the temperature continues to drop. A positive surface density profile for the condensed ice has been developed from ~4 to 20 $R_U$. The condensed silicate also follows a similar surface density profile because it is able to stick onto the condensed ice. This is ideal for generating a satellite system with bulk composition close to 1 to 1 in its rock-to-ice ratio.

Moving from the outer to the inner region of the disc, the silicate surface density starts deviating from the ice surface density profile at ~7 $R_U$ and eventually falls to zero at ~4 $R_U$ at $t_{disc}$ = 10,000 years. This leaves the current Miranda region more ice dominated. This is caused by the non-negligible gas drag acting on the micrometre size silicate particles. Hence, some of the condensed silicate particles are lost to Uranus due to the gas drag induced radial drift. We show in Section 3.3 that this is ideal for explaining the more icy composition and the lower mass of the innermost regular satellite Miranda.

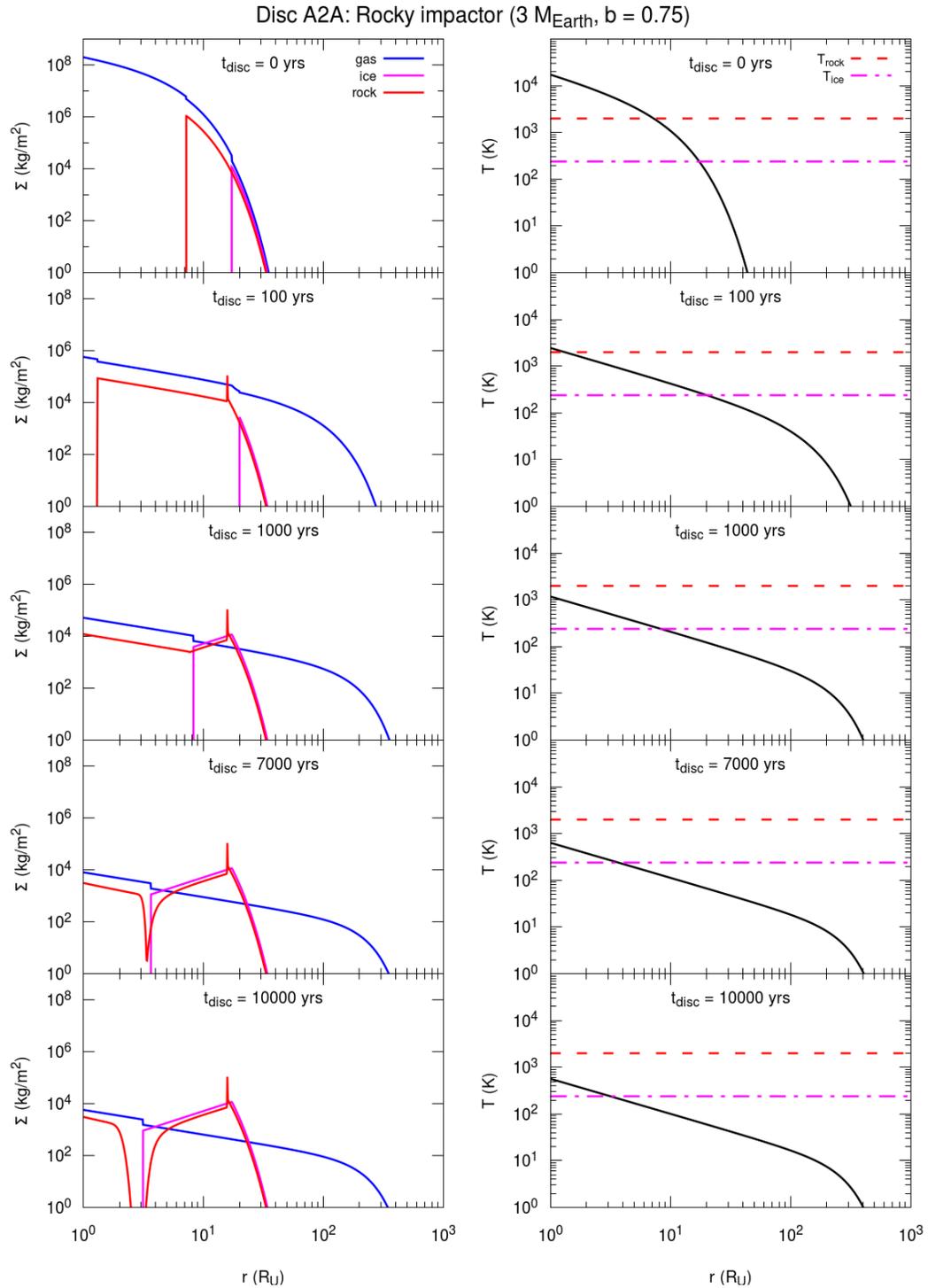

Fig. 2 - The left panel shows the evolution of the surface density of the gaseous component (blue), condensed ice (magenta) and condensed silicate (red) of the impact generated disc from $t_{disc} = 0$ to 10,000 years. The disc is generated by a fully rocky impactor colliding with proto-Uranus with $b$ = 0.75 (Disc A2A of Table 1). The right panels depict the temperature profile of the disc, compared to the silicate condensation temperature (2000 K; red dashed line) and the water ice

condensation temperature (240 K; magenta dashed-dotted line). At the end, the total condensed solids in the orbits (6.3 × 10$^{-5}$ $M_U$ of ice + 5.5 × 10$^{-5}$ $M_U$ of silicate) follow a surface density profile with a positive gradient from ~4 to 20 $R_U$ and the inner region (4 $R_U$ < $r$ < 7 $R_U$) is dominated by ice.

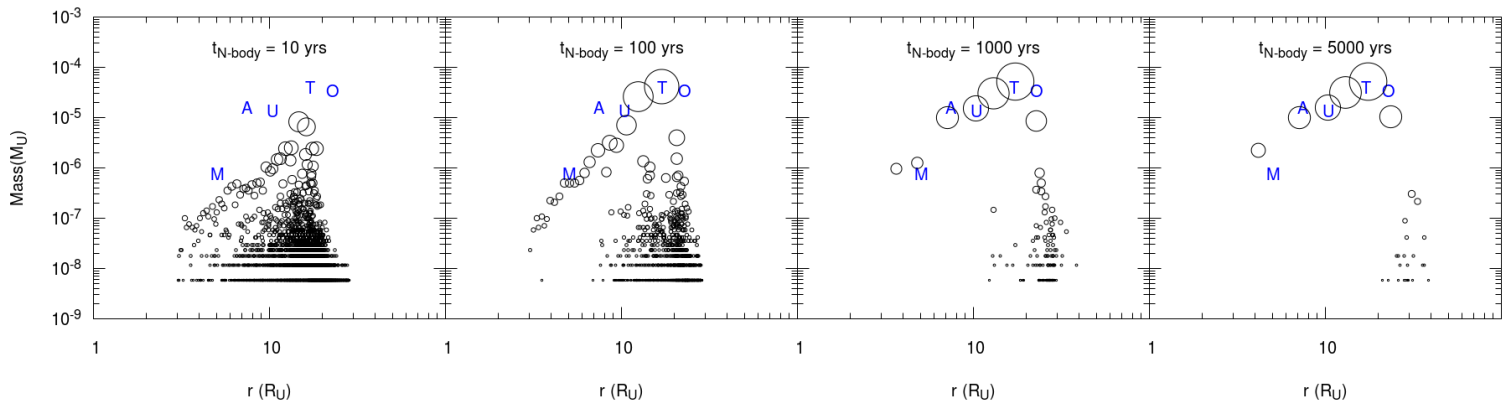

Fig. 3 - Growth of the satellites from a disc of equal mass moonlets in the N-body simulations from $t_{N-body}$ = 10 to 5000 years. The initial condition of the moonlets follows the condensed solid surface density presented at $t_{disc}$ = 10,000 years in Fig. 2 (Disc A2A of Table 1). The blue capital letters represent the current mass of the five regular Uranian satellites. The satellites we form, which have a total mass of ~10$^{-4}$ $M_U$, follow a similar mass-distance distribution with the current system.

### 3.2. Growth of the satellites

Next, we generate a disc of moonlets based on the surface density profile of condensed solids at $t_{disc}$ = 10,000 years as in Fig. 2 and perform the N-body simulation for 5,000 years to form a satellite system from the disc (as described in Section 2.3). Fig. 3 shows the mass evolution of the satellite system starting from a disc of equal mass moonlets. $t_{N-body}$ = 0 years represents the formation time of the moonlet. As in Ishizawa et al. (2019) and Ida et al. (2020), we assume the moonlets form at the same time throughout the whole disc for simplicity. In reality, the outer region could begin forming moonlets earlier since it reaches a high solid to gas surface density earlier (see Fig. 2).

Due to a higher solid surface density in the outer region of the disc, more distant satellites grow more quickly and to larger size. Umbriel-sized objects (~10$^{-5}$ $M_U$) form within 10 years (~2,500 orbits of Miranda) of the formation of the moonlets at ~20 $R_U$. Satellites with the mass of Titania (the largest satellite of Uranus) emerge within 100 years. At $t_{N-body}$ = 1,000 years, nearly all the masses within ~10 $R_U$ are accreted by the Miranda, Ariel, and Umbriel analogues, while the outer region is still crowded with moonlets with masses less than Miranda's mass. These leftover moonlets are the residual of a decreased solid surface density beyond 20 $R_U$ in Fig. 2, and they are slowly cleared by or accreted by the two outermost satellites in the current Titania and Oberon region from 1,000 to 5,000 years.

This particular case not only has the potential to reproduce the tilt and internal heat flux of Uranus, but also likely results in a satellite system comparable to the current system, in terms of the mass-distance distribution as well as the total number of the satellites. Similar to results of Ida et al. (2020), our simulation reproduces the current mass-distance distribution of the system, with the satellites' masses in general increasing with semi-major axis.

### 3.3. Bulk composition of the final system

In contrast to the previous studies on the formation of Uranus satellites (Ida et al., 2020; Ishizawa et al., 2019), we also consider the rocky component of the disc. Hence, we can study which impact cases lead to a satellite system with bulk compositions (i.e density) comparable to the current satellites. Our results indicate that a rocky impactor is required to reproduce the ~1 to 1 rock-to-ice ratio of the current system. After the giant impact, part of the target's and impactor's materials are ejected to form a circumplanetary disc. Since the mantle of Uranus is assumed to be water-rich (Helled et al., 2011; Nettelmann et al., 2013) in our SPH simulations, the impactor has to be rock-dominated in order to form a satellite system with a mixture of rock and ice. A large impact parameter collision ($b ≥ 0.6$) is also required to generate Uranus tilt, which is a collision that cannot dredge up the silicate material deep in Uranus interior (if there are any).

Fig. 4 shows the final uncompressed density of the satellite system computed from our N-body simulation. Similar to the mass-distance distribution, there is a positive trend for the uncompressed density of the satellites, i.e. satellite density in general increases with orbital distances. According to Fig. 2, after $t_{disc}$ = 1,000 years part of the silicate condensed in the inner region (< 7 $R_U$) are lost to Uranus through gas drag, while the outer region remains a constant ~1 to 1 condensed rock-to-ice ratio. This is because the silicate can stick onto the condensed ice and stay in their orbit in the outer region, whereas ice has not yet condensed in the inner region and silicates which feel the gas drag drift inward. As a result, the inner disc is more icy than the outer disc.

Based on the Voyager 2 measurements on masses and radii (Jacobson et al., 1992), the current satellites also show such a positive trend in their mean density. The blue data points in Fig. 4 represent the measured mean density of the satellites. The innermost satellite Miranda has the lowest mean density among the satellites, while the outermost large satellites Titania and Oberon could contain a higher mass fraction of rock than Miranda. Our model with condensed silicate loss through gas drag in the inner disc could naturally explain the lower density of Miranda. Nevertheless, we emphasize that more accurate measurements on masses and radii of the Uranian satellites are required to confirm whether this density trend is robust.

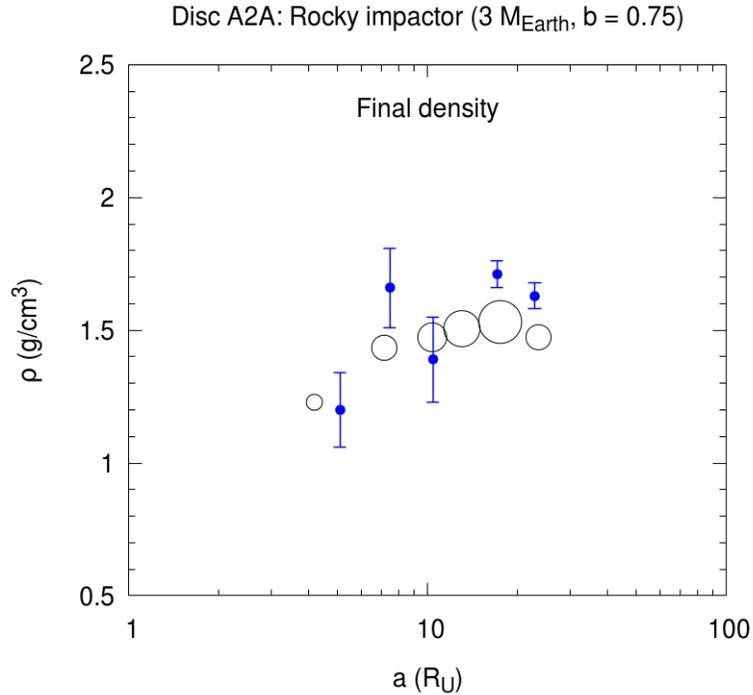

Fig. 4 - Uncompressed density of the final satellites at $t_{\text{N-body}} = 5{,}000$ years from the N-body simulations shown in Fig. 3 (Disc A2A of Table 1). The blue points and error bars represent the current density of the Uranus satellites taken from Jacobson et al. (1992). Only densities for objects with masses larger than $6 \times 10^{-7}$ $M_U$ are shown. Note that our simulation reproduces the apparent positive trend of density, with the satellites' densities in general increasing with orbital distances in general.

## 4. Investigation of different parameters

We also study the impact results other than those presented in the previous section, for example by changing the alpha viscosity parameter, assuming an icy impactor, changing the solid-gas interaction, as well as including solid diffusion in our model.

### 4.1. The sensitivity of the results to the disc's $\alpha$ parameter

In our most successful case, the viscosity parameter is set to $\alpha = 10^{-3}$. The exact value of $\alpha$ of a circumplanetary disc or impact generated disc is highly uncertain. It has been suggested that $\alpha$ should range from $10^{-2}$ to $10^{-5}$ (e.g. Birnstiel et al., 2010). Fig. 5 shows the effect of changing $\alpha$ to the surface density of the gaseous component and the condensed solid. When decreasing $\alpha$ to $10^{-5}$ (right panels of Fig. 5), viscous heating becomes less effective and thus lead to a lower temperature in the disc (since $T \sim \alpha^{1/4}$) (Hartmann et al., 1998). This results in a higher condensed rock and ice surface density in the low-$\alpha$ disc throughout the disc evolution. Hence, the satellite system to form in a low-$\alpha$ disc is likely to be too massive, compared to our successful case presented in Section 3. On the other hand, increasing $\alpha$ to $10^{-2}$ likely results in a satellite system with too low mass, as less rock and ice condensed due to stronger viscous heating of the disc.

Another effect of changing $\alpha$ is that the diffusion timescale $t_{diff}$ of the disc is also varied, as $t_{diff} \sim \alpha^{-1}$. A low-$\alpha$ disc spreads much slower than a high-$\alpha$ disc because of the lower efficiency of angular momentum transport from the inner disc to the outer disc when the viscosity is lower. As a result, the gas (H-He) surface density of the low-$\alpha$ disc remains a few orders of magnitude higher than that of the high $\alpha$ disc at $t_{disc}$ = 10,000 years. Furthermore, at the end of simulation only regions at ~10 $R_U$ in the low-$\alpha$ disc achieve a high solid-to-gas ratio (> 1) for moonlet formation to take place. Thus a satellite system is likely to emerge slower in a low-$\alpha$ disc. In addition, the slower spread of the disc aided the formation of a massive satellite system, as less silicate and water ice remain in a gaseous state and being lost into Uranus during the viscous spreading of the disc.

We demonstrated that the choice of $\alpha$ could be crucial in forming the final system with the correct mass, with the $\alpha$ = $10^{-3}$ in Section 3 yielding the best results. Although both the higher- and lower- $\alpha$ discs are less successful in forming the current Uranian satellite system, we cannot completely rule out the former case. We expect the effect of gas drag and Type-I migration on the formed moonlets to be more significant in the low-$\alpha$ disc as the gas surface density remains high for a much longer period of time. Strong gas drag and Type-I migration could potentially lead to loss of moonlets or even fully formed satellites, thus lowering the total mass of the satellite system. Future investigations including both gas effects are required to verify this.

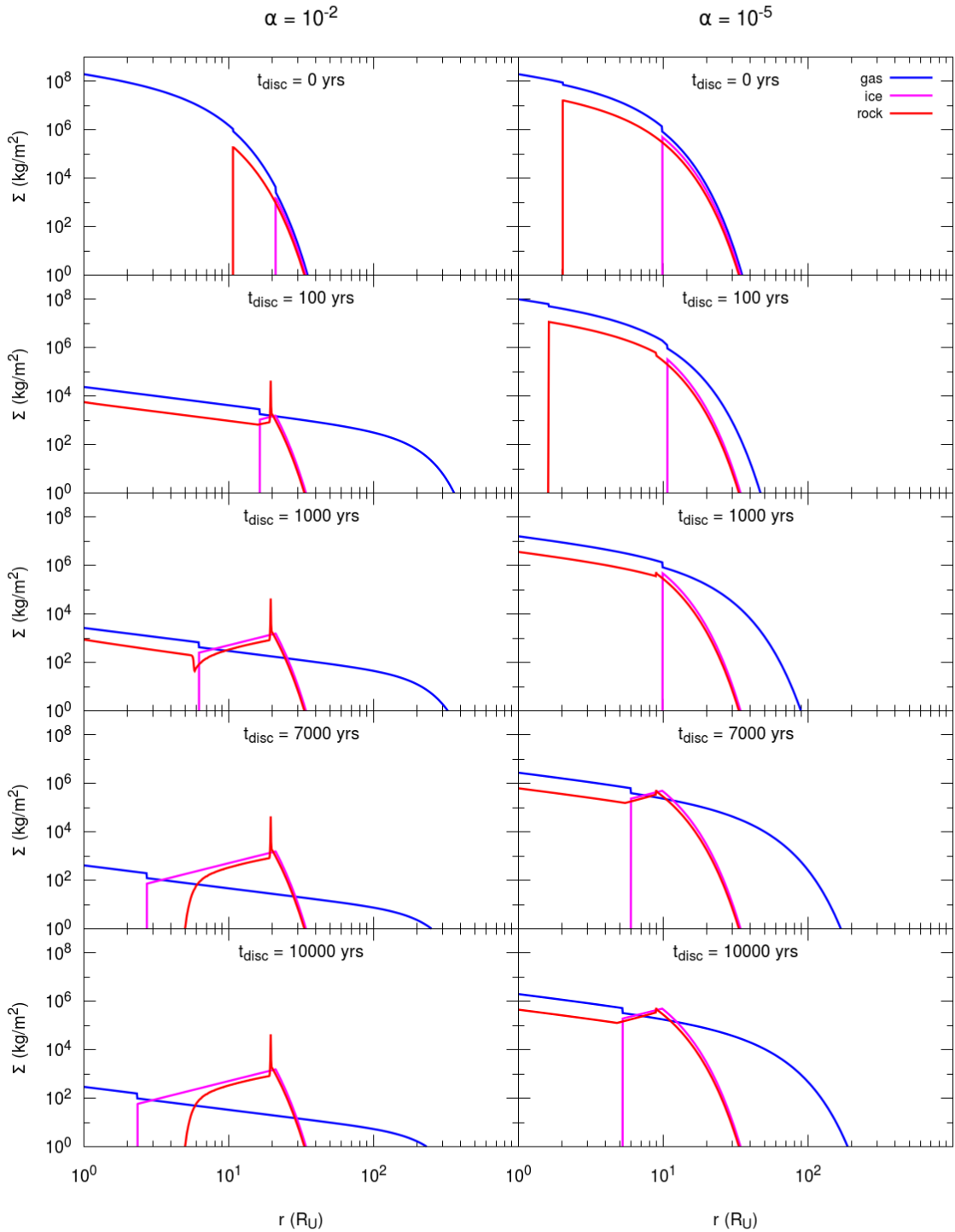

Fig. 5 - Same as the left panels of Fig. 2, but assumes a higher $\alpha = 10^{-2}$ (left panels) and a lower $\alpha = 10^{-5}$ (right panels) for the disc, respectively.

### 4.2. Icy impactor results

Assuming an icier impactor colliding with Uranus would lead to formation of a too icy satellite system. Fig. 6 shows the disc evolution and the N-body simulation results after a 3 $M_{Earth}$ impactor with 50% ice and 50% rock colliding with the proto-Uranus with $b = 0.6$ (Disc CA of Table 1). Similar to the results of a rocky impactor, after 10,000 years of disc spreading the condensed solid adopted a positive gradient of surface density profile from ~3 to 20 $R_U$. However, different from Fig. 1, the solid surface density of condensed ice is roughly two orders of magnitude higher than the condensed silicate. This is because rock is enveloped by the icy mantle in the impactor, and thus the mass of rock ejected to the disc is two orders of magnitude lower in this case (Table 1). Even though the final N-body result follows a mass-distance distribution similar to the current system, the final system is too water-rich, with the satellites all having uncompressed density close to 1 g/cm$^3$. Hence, compared to results presented in Section 3 for rocky impactor, results for an icy impactor are less successful. However, as discussed in Section 4.2., there are uncertainties regarding the internal composition of Uranus. If Uranus contains a higher fraction of rock (Helled et al., 2011; Teanby et al., 2020), it allows a more icy impactor to form the current satellites with their documented half-rock-half-ice composition.

Due to the limit of computational resources, we are only able to perform SPH simulation with $b = 0.6$ for the icy impactor. With a slightly larger $b$, similar to the successful case we presented in Section 3 ($b = 0.75$), we expect more rock could be ejected to disc. However, it is difficult to overcome a two order of magnitude difference between rock and ice mass by increasing $b$ only. Hence, we expect the over-icy composition of the satellites obtained in this case would still persist even if the impact is slightly more oblique.

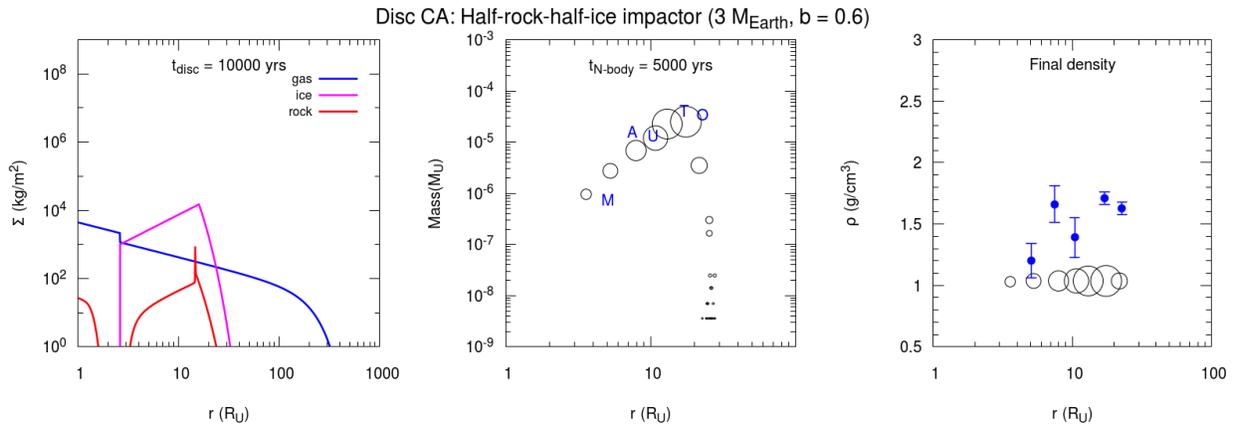

Fig. 6 - Results of the disc evolution (left panel), masses (middle panel) and the uncompressed density (right panel) of the final satellites from the N-body simulation. The disc is generated by a half-rock-half-ice impactor with a slightly lower impact parameter $b = 0.6$ (Disc CA of Table 1). See Fig. 2, Fig. 3 and Fig. 4 for figures notation and colour description.

### 4.3. Changing the solid-gas interaction

The assumptions of the solid-gas interaction we made in Section 3 (point (2) to (4)) are also important in determining the final system. Fig. 7 depicts results for the same rocky impactor as in our successful case in Section 3, but assuming both condensed silicate and ice stay at micrometre size and silicates do not stick onto ice throughout the whole disc evolution. Condensed solids at micrometre size couple well with the gas ($St \rightarrow 0$) and follow a solid surface density similar to gas surface density with negative gradient, although both rock and ice drift inward due to gas drag. A solid surface density with negative gradient has difficulty in forming Uranian satellites with their current mass-distance distribution since more condensed mass is concentrated in the region within 10 $R_U$. Our N-body results indicate that instead of a few large satellites with masses in between $10^{-5}$ to $10^{-4}$ $M_U$, we form more than 20 small satellites with masses roughly $10^{-7}$ to $10^{-6}$ $M_U$. The total mass of the system is also an order of magnitude less massive than the current system, even though most of the satellites have a half-rock-half-ice composition. Hence, we conclude that the assumptions of the solid-gas interaction for rock and ice in Section 3 is crucial to successfully match the mass-distance distribution and bulk composition of the current Uranian satellite system.

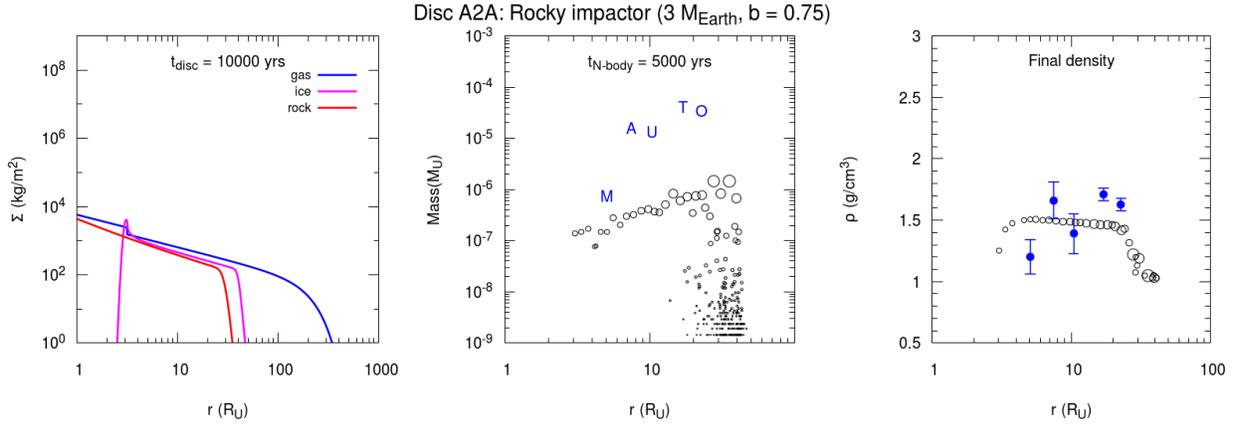

Fig. 7 - Same as Fig. 6, but assuming both condensed silicate and ice remain micrometre size ($St \rightarrow 0$) and evolve separately (i.e. silicate not sticking to ice) during the spreading of the disc. The impactor and its impact parameter is the same as the rocky impactor in our successful case in Section 3.

### 4.4. Including diffusion of solid

We also run the semi-analytical 1D model described in Section 2.2 adding turbulence-driven diffusion to it. The equation that has been solved for rock and ice density evolution in this case is the following (Drążkowska and Alibert, 2017)

$$\frac{\partial \Sigma_s}{\partial t} + \frac{1}{r}\frac{\partial}{\partial r}(\Sigma_s v_r r) = \frac{1}{r}\frac{\partial}{\partial r}\left[D \Sigma_s r \frac{\partial}{\partial r}\left(\frac{\Sigma_s}{\Sigma_g}\right)\right]$$

(8)

where $\Sigma_s$ is the surface density of condensed silicate or ice, $D$ is the solid diffusion coefficient, $\Sigma_g$ is the gas density and $v_r$ is the same as equation Eq. 6 in Section 2.2. The solid diffusivity can be estimated as $D = \frac{\nu}{1+St^2}$ (Birnstiel et al., 2010; Youdin and Lithwick, 2007) where $St$ is the Stokes number and $\nu$ is the viscosity (same as Section 2.2). Consequently in our model, the diffusivity of condensed rock will be $D \approx \nu$ as $St \to 0$ for rcok and the diffusivity of condensed ice will be $D \approx 0$ as $St \to \infty$ for ice (see Section 3).

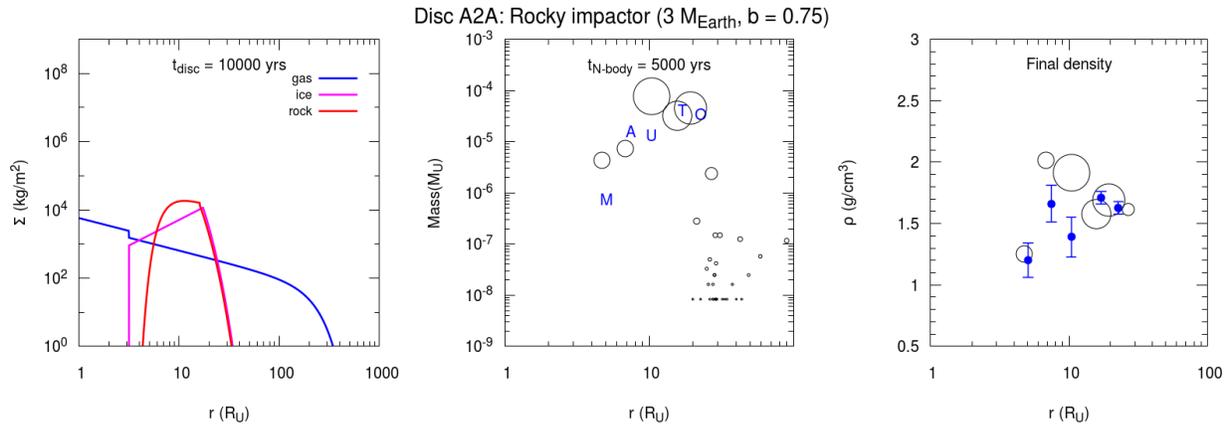

Fig. 8 - Same as Fig. 6, but including turbulence-driven diffusion for the condensed solid. The impactor and its impact parameter is the same as the rocky impactor in our successful case in Section 3.

We found that including diffusion of condensed solid does not affect our final results significantly. Fig. 8 shows the results of the disc evolution and the N-body simulations when diffusion of condensed solid is included. Compared the left panel of Fig. 8 to Fig. 2, the ice density distribution does not change at all, since condensed ice do not suffer any diffusion. The silicate surface density drops to zero in the innermost part of the disc within 3 $R_U$, where diffusion is dominant but satellites do not form there since this region is within the Roche limit. In the outer part of the disc, where satellites actually form and evolve, the solid distribution does not change drastically. In fact, the middle and right panels of Fig. 8 show similar satellites' configurations and density when compared to Fig. 3. Even if the higher silicate density in the outer part led to formation of larger and more rocker satellites in the region of Ariel and Umbriel, the general positive trend of the mass and density of the satellites is maintained. Hence, the effects of diffusion can be neglected in terms of the final satellite configuration.

### 5. Discussion

#### 5.1. Uncertainties in Uranus' composition and its history

The only spacecraft that has visited Uranus is *Voyager 2* in 1986. Due to the limited data our understanding of Uranus is very incomplete and there are still many uncertainties regarding this planet, including its exact bulk composition and internal structure (see reviews by Helled et al., 2020 and; Helled and Fortney, 2020 for discussion). It is often assumed, as in this study, that Uranus is mainly composed of water ice (~90 % of mass; Helled et al., 2011; Nettelmann et al., 2013), however, the actual abundance of water ice in Uranus is far from certain. A composition including ~82% of rock with the rest made of H-He can also match its observed properties (e.g.

density, gravity field; Helled et al., 2011), although this is an extreme and improbable case. If Uranus contains a much higher amount of rock (e.g. ~50 % of mass) than most models have assumed, which can indeed be the case (Teanby et al., 2020), then the requirement of a rocky impact to generate the current satellite system could be relaxed. In such a case, the rocky component of the satellites can then also be delivered by the target of the collision.

We aim to reproduce the features of the current satellite system from our models. However, whether the Uranian system stays the same since its formation is still under debate. As suggested by the *Nice* model (Tsiganis et al., 2005), Uranus and Neptune could have undergone a phase of instability and inject huge amounts of Kuiper belt material into the inner solar system (e.g., Gomes et al., 2005; Mojzsis et al., 2019). It has been shown that this injected Kuiper belt material could also bombard the Uranian satellites, causing loss of volatiles, mass erosion and even catastrophic destruction of satellites (Nimmo and Korycansky, 2012; Wong et al., 2019). Hence, the Uranian system, prior to giant planet instability, could be more massive and ice rich. If that is the case, it would favor an impactor more enriched in ice (e.g. results in Section 4.2).

### 5.2. Effect of tidal evolution

Given the short N-body simulation time, we neglect the tidal effect from Uranus on the satellite system. In principle, all satellites that form beyond the synchronous radius (~ 3 $R_U$) of Uranus would migrate outward through tidal interaction with Uranus and thus tidal evolution may play a role in shaping the final satellite system after its formation. However due to the uncertainty in Uranus' internal structure, the tidal quality factor $Q_U$ of Uranus, which determines the migration rate of the satellites, is poorly constrained. Nevertheless, according to the dynamical study of Tittemore & Wisdom (1990), the $Q_U$ of Uranus should not be less than 11,000. Otherwise, Ariel and Umbriel would have captured into their 2:1 mean-motion resonance (MMR), which is difficult to escape. Currently none of the satellites of Uranus are trapped in any low order MMR (Cuk et al. 2020). Adopting the following tidal evolution equation (Charnoz et al. 2010) for satellites with circular orbits beyond the corotation radius,

$$\frac{da}{dt} = \frac{3k_{2U}MG^{1/2}R_U^5}{Q_U M_U^{1/2} a^{11/2}},$$

(9)

we find that nearly all satellites formed in our successful system shown in Fig. 4 migrated tidally outward by less than 0.5 $R_U$ within 4.5 Gyr; assuming $Q_U$ = 11000 and $k_{2U}$ = 0.104 (Gavrilov & Zharkov 1977), where $k_{2U}$ is the tidal Love number of Uranus and $M$ is the satellite mass. Even if we assume $Q_U$ has a small value (a few thousands) similar to recently measured for Saturn (e.g. Lainey et al., 2017), most of the satellites migrated outward only by less than 1 $R_U$ within the lifetime of the solar system. We therefore conclude that our results are relatively insensitive to the effect of tidal evolution of the satellite system.

### 5.3. Rocky world beyond the ice line

Our results suggest that rocky objects could dominate the region beyond the ice line. We expect ice (including $H_2O$, $CH_4$, $NH_3$) to rock condensate in a ratio of ~2 to 1 for solar composition gas (Lodders, 2003). However, some of the Kuiper objects have a rocky dominated composition. For example one of the largest Kuiper belt objects, Pluto, is likely to be 70% rock (McKinnon et al., 2017). Triton, which is the largest satellite of Neptune and likely captured from the Kuiper belt (Agnor and Hamilton, 2006; McCord, 1966; McKinnon, 1984), also consists of more than 70% of rock (McKinnon and Kirk, 2014). Combined with our results showing that a ~3 $M_{Earth}$ rock dominated impactor is required to form the satellites of Uranus, as well as the possibility of a more rocky Uranus (see previous subsection), all the evidences suggested that we may have to revise our picture of the formation of the outer solar system and gain a better understanding of how it could acquire such a rocky composition beyond the ice line.

## 6. Conclusions

The giant impact scenario remains the standard scenario to explain Uranus' properties (Kegerreis et al., 2018; Kurosaki and Inutsuka, 2018; Reinhardt et al., 2020) and the formation of its satellite system (Ida et al., 2020; Ishizawa et al., 2019). In this study, we revisited the SPH simulation results of the successful simulations of R20 to study which impact scenario best reproduces the satellite system of Uranus. By including new sets of SPH simulations with more advanced equation of state (Thompson and Lauson, 1974; Becker et al., 2014), followed by evolving the impact generated disc in a 1D disc evolution model (Cilibrasi et al., 2018, 2021), and growing the satellites from a disc of moonlets with N-body simulations (Grimm and Stadel, 2014), we found that the mass-distance distribution and the bulk composition of the satellites are best reproduced with a rocky impactor (100% rock) colliding with Uranus at $b = 0.75$ and $v_{imp} = 18.2$ kms$^{-1}$. Such an oblique collision is also likely to explain the large tilt and the low internal heat flux of Uranus (R20). We also found that the assumption of solid properties are important. In order to reproduce the positive trend of the mass-distance distribution and the similar trend of the mean density of the system, condensed icy particles should grow quickly to decouple from the gas ($St \rightarrow \infty$), whereas the silicate particles should remain small ($St \rightarrow 0$) unless sticking to the condensed ice in the same orbit (Ida et al., 2020). The giant impact scenario can naturally explain the key features of Uranus and its regular moons. We therefore suggest that the Uranian satellite system formed as a result of an impact rather than from a circumplanetary disc

While we can successfully reproduce the Uranian system, our model makes simplifying assumptions on the solid properties and neglects some detailed physical processes involved in solid condensation. For instance, we assume constant condensation temperatures for rock (2000 K) and ice (240 K). In reality, the condensation temperature of a material depends on the internal pressure of the gas, and the pressure changes when the disc spreads. We also neglect the release of latent heat when the solid condenses and this may change the temperature and pressure profile of the disc. Future studies should include detailed dust physics in studying the viscous spreading of the disc.

Our results also suggest that rocky objects could dominate the region beyond the ice line. However, we neglected several uncertainties regarding Uranus and its satellite system, including whether Uranus could have a more rocky bulk composition (Helled et al., 2011; Teanby et al., 2020) and whether the current Uranian system has remained unchanged since formation. The proposed *Nice* model instability of the giant planets (Tsiganis et al., 2005) could have a decisive impact on shaping the Uranian satellite system after formation (Nimmo and Korycansky, 2012; Wong et al., 2019). Future studies on Uranian satellite formation should also consider these uncertainties. Finally, we suggest that accurate measurements of the compositions, the tidal quality factor $Q_U$ of Uranus and its regular moons with a future space mission (e.g. Fletcher et al., 2020) could constrain the origin and early evolution of Uranus and its satellite system. This in return will also shed light on the origin of our planetary system.

**Acknowledgments**

We would like to thank Thomas Meier and Jens Oppliger for their valuable contributions in implementing ANEOS and REOS3 for the SPH simulations presented in this paper. We would also like to thank Lucio Mayer, Judit Szulágyi for helpful discussions regarding the long term evolution of the proto-satellite disc. We thank Sebastien Charnoz and the other referee in reviewing the manuscript and giving suggestions to improve it. CR and RH thank Andreas Becker, Nadine Nettelmann, Yamila Miguel and Roland Redmer for fruitful discussions regarding the REOS3 equation of state. This work has been carried out within the framework of the National Center of Competence in Research PlanetS, supported by the Swiss National Science Foundation (SNSF). The authors acknowledge the financial support of the SNSF. The authors acknowledge the computational support from Service and Support for Science IT (S$^3$IT) of University of Zurich and the Swiss National Supercomputing Centre (CSCS). RH acknowledges support from the Swiss National Science Foundation (SNSF) via grant 200020_188460.

**Appendix A1. Disc parameters**

We study the impact simulations from R20 that are most favorable in explaining Uranus' features and have the potential of forming the Uranian satellites based on the criteria set by R20. In addition, we perform extra sets of SPH simulation with the more advanced AENOS / REOS3 EOS. Table 1 shows the parameters of the impact generated disc from the SPH simulations that we investigated. In general, the impact generated disc has a total rock and ice mass at least an order of magnitude higher than the current satellite system. As we have shown in Section 3, both rock and ice are lost to Uranus through viscous spreading, causing less than 1% of rock and ice in the original disc to condense and eventually forming the moonlets of the system. Hence, an impactor of about 3 $M_{Earth}$ (0.206 $M_U$) is required to form the satellite system with sufficient mass.

Table 1 - Parameters of the impact generated disc at the end of the SPH simulations (i.e. before vicious spreading). We model the proto-Uranus and the impactor with the Tillotson / ideal gas and ANEOS / REOS3 equation of state, respectively. In this table, we show the impactors' mass ($M_{imp}$), impactors' composition, impact parameter ($b$) and impact velocity ($v_{imp}$) of the collision, total mass of the impact generated disc ($M_{disc}$), total rock mass ($M_{disc,rock}$) and ice mass ($M_{disc,ice}$) in the disc, total angular momentum of the disc ($J_{disc}$) and the mean radius of the disc ($r_{disc}$). We highlight the most successful disc (A2A) in generating the current features of the Uranian satellite system.

| | Disc | $M_{imp}$ ($M_U$) | Impactor composition | $b$ | $v_{imp}$ (km/s) | $M_{disc}$ ($10^{-3} M_U$) | $M_{disc,rock}$ ($10^{-3} M_U$) | $M_{disc,ice}$ ($10^{-3} M_U$) | $J_{disc}$ ($10^{15}$cm$^2$s$^{-1}$) | $r_{disc}$ ($R_U$) |
|---|---|---|---|---|---|---|---|---|---|---|
| Tillotson/ Ideal gas | A1T | 0.138 | 100% rock | 0.75 | 20.45 | 12.8 | 9.62 | 1.66 | 6.47 | 3.03 |
| | A2T | 0.206 | | | 20.53 | 44.8 | 22.6 | 8.46 | 6.01 | 2.61 |
| | B1T | 0.138 | 12% rock; 88% ice | 0.65 | 19.48 | 7.28 | 0.03 | 6.49 | 6.98 | 3.52 |
| | B2T | 0.206 | | 0.60 | 19.51 | 26.9 | 0.23 | 18.8 | 5.99 | 2.60 |
| AENOS/ REOS3 | A1A | 0.138 | **100% rock** | **0.75** | 18.47 | 10.3 | 4.61 | 2.59 | 6.36 | 2.93 |
| | **A2A** | **0.206** | | | **18.20** | **49.9** | **9.21** | **15.1** | **5.94** | **2.55** |
| | B1A | 0.138 | 12% rock; 88% ice | 0.65 | 17.90 | 2.89 | 0.00 | 2.06 | 6.34 | 2.91 |
| | B2A | 0.206 | | 0.60 | 17.60 | 25.5 | 0.01 | 13.6 | 5.84 | 2.47 |
| | CA | 0.206 | 50% rock; 50% ice | 0.60 | 17.81 | 26.4 | 0.06 | 12.2 | 5.82 | 2.45 |

| | DA | | 27% rock; 73% ice | | 17.68 | 26.1 | 0.08 | 13.5 | 5.83 | 2.46 |
|---|---|---|---|---|---|---|---|---|---|---|
| | EA | | 73% rock; 27% ice | | 17.96 | 24.9 | 0.32 | 8.35 | 5.82 | 2.45 |

**Appendix A2. The importance of a physically realistic EOS**

In Section 3, we apply the AENOS / REOS3 EOS to model the internal structure of proto-Uranus and the impactor. We found that this yields slightly different results compared to the Tillotson / ideal gas models that R20 adopted. Fig. 9 shows the results of the disc evolution (left panel), mass (middle panel) and the uncompressed density (right panel) of the final satellites from the N-body simulation for the same impactor as descripted in Section 3, but with Uranus and the impactor modelled by the Tillotson / ideal gas EOS. Generally, we find that the total bound mass as well as the disc mass agree very well for both EOS combinations (see Table 1). However, in the case of the Tillotson / ideal gas EOS more rock originating from the impactor and less ice from proto-Uranus' inner envelope is ejected into orbit compared to the corresponding ANEOS / REOS3 simulations. This is the result of the rocky impactor being more compact in case of ANEOS, likely due to a more accurate treatment of the high pressure behaviour of rocks, which means that it is more difficult to be tidally disrupted and therefore less rock is deposited in the disk in the impact. The final satellite system generated by the Tillotson model is slightly more massive (> $2 \times 10^{-4}$ $M_U$; more than two times the current system) and is more rocky (ice to rock ratio ~ 0.3) than results of the AENOS model, even though the overall positive trend of the mass-distance distribution and the bulk density can still be reproduced.

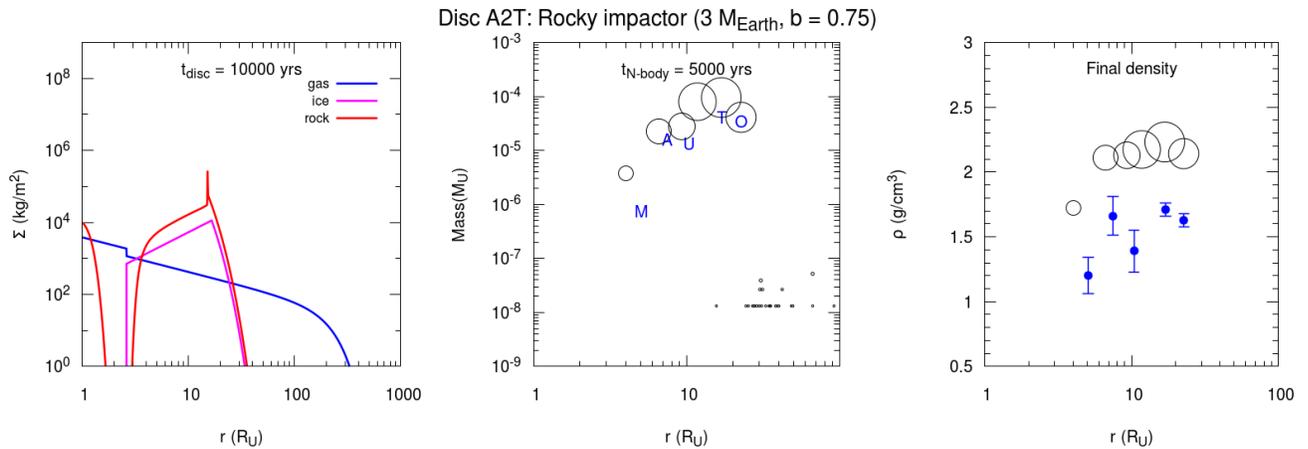

Fig. 9 - Results of the disc evolution (left panel), masses (middle panel) and the uncompressed density (right panel) of the final satellites from the N-body simulation. The disc is generated by the same impactor with the same *b* as the results shown in Section 3, but the proto-Uranus' and the impactors internal structure is described by the Tillotson / ideal gas EOS (Disc A2T of Table 1). See Fig. 2, Fig. 3 and Fig. 4 for figures notation and colour description.